\documentclass[manuscript]{acmart}

\AtBeginDocument{%
  }

\usepackage{subcaption}
 \usepackage[ruled,vlined]{algorithm2e}

\begin{document}


\title{Exploring Dynamic Load Balancing Algorithms for Block-Structured Mesh-and-Particle Simulations in AMReX}

\author{Amitash Nanda}
\authornote{Both authors contributed equally to this research.}
\orcid{0000-0002-9792-2110}
\affiliation{%
  \institution{University of California San Diego}
  \city{San Diego}
  \state{California}
  \country{USA}
}
\email{ananda@ucsd.edu}

\author{Md Kamal Hossain Chowdhury}
\orcid{0000-0002-7191-3240}
\authornotemark[1]
\affiliation{%
  \institution{The University of Alabama}
  \city{Tuscaloosa}
  \state{Alabama}
  \country{USA}}
\email{mhchowdhury@crimson.ua.edu}

\author{Hannah Ross}
\orcid{0000-0001-9646-793X}
\affiliation{%
  \institution{Barcelona Supercomputing Center}
  \city{Barcelona}
  \country{Spain}
  }
  \email{hannah.ross@bsc.es}

\author{Kevin Gott}
\orcid{0000-0003-3244-5525}
\affiliation{%
 \institution{Lawrence Berkeley National Lab}
 \city{Berkeley}
 \state{California}
 \country{USA}}
\email{kngott@lbl.gov}

\renewcommand{\shortauthors}{Nanda, Chowdhury, Ross, Gott}

\begin{abstract}
Load balancing is critical for successful large-scale high-performance computing (HPC) simulations. With modern supercomputers increasing in complexity and variability, dynamic load balancing is becoming more critical to use computational resources efficiently. In this study, performed during a summer collaboration at Lawrence Berkeley National Laboratory, we investigate various standard dynamic load-balancing algorithms. This includes the time evaluation of a brute-force solve for application in algorithmic evaluation, as well as quality and time evaluations of the Knapsack algorithm, an SFC algorithm, and two novel algorithms: a painter's partition-based SFC algorithm and a combination Knapsack+SFC methodology-based on hardware topology. The results suggest Knapsack and painter's partition-based algorithms should be among the first algorithms evaluated by HPC codes for cases with limited weight deviation and will perform at least slightly better than AMReX's percentage-tracking partitioning strategy across most simulations, although effects diminish as weight variety increases.
\end{abstract}




\maketitle

\section{Introduction}
High-performance computing (HPC) is an essential tool in modern scientific research and has a wide range of applications, such as meta-genomics, climate modeling, molecular dynamics, and astrophysics \cite{abud2023highly, pavlopoulos2023unraveling, de2023double, nguyen2023high, balaguru2023increased, hackett2024gravitational}. The operation of large-scale, decomposable scientific simulations on HPC supercomputers often depends on effective dynamic load balancing \cite{tarraf2024malleability, rathore2024load, ilsche2024optimizing}.

Uneven workloads can lead to significant resource under-utilization and restrict the feasible HPC simulation space. Applications like fluid dynamics, molecular modeling, and climate modeling rely on effective load balancing to prevent bottlenecks and ensure both new and repeated simulations run productively \cite{micale2024increasing, jakobs2024parallelization}. However, achieving an optimal balance is challenging for various reasons, including difficulty capturing or defining an accurate weighting factor, limitations of the approximation algorithm, and variation of computational loads due to the problem's complexity or evolution over time \cite{buch2023vector, gopalakrishnan2016adaptive,clarke2011dynamic}. Recent advances in distributed AI optimization and decentralized model training in heterogeneous environments provide valuable insights for developing dynamic load-balancing algorithms in HPC that can effectively manage trade-offs between compute, memory, and communication resources \cite{nanda2024cptquant,balija2024building}. 

In this research, performed as part of NERSC's summer internship program \cite{NERSCsummer}, we investigate the current state of load balancing algorithms in AMReX, a block-structured adaptive mesh refinement (AMR) software framework \cite{AMReX_JOSS}. AMReX's current load-balancing algorithms are a Knapsack algorithm and a Morton space-filling curve (SFC) \cite{AMReX_JOSS, zhang2021amrex, rowan2021situ}. Our work expands on these algorithms by investigating the parallelization of a brute force algorithm, a novel hybrid load-balancing algorithm combining SFC and knapsack, and an improved SFC bisection strategy using the painter's algorithm \cite{PPPGeeks, PPPTakeUForward, PPPApna}. We statistically compare the efficiency and timing to identify the best use cases of these algorithms.

\section{Load Balancing Algorithms}
When parallelizing an application, it is essential to distribute the workload evenly among processors to ensure the total run time of the processes is minimal. Although this concept seems straightforward, its implementation can be one of the most complex and impactful issues in large-scale HPC simulations \cite{jin2011high}. In addition to knowing how much work each of these boxes will perform, limits on the collection of useful weighting data, the available memory on each processor and the overhead from redistributing the boxes must also be considered. As a first step to explore potential improvements to load balancing design and implementation, this research focuses on algorithmic performance to inform algorithm selection and design. The additional complexities with integrating chosen algorithms will be studied in future work.

In AMReX applications \cite{zhang2021amrex}, the domain is divided into rectilinear `boxes' of cells, which are spread between MPI ranks. Let \( N \) boxes, each with a weight \( w_1, w_2, \dots, w_j, \dots, w_N \), be assigned to \( P \) processors. The load on processor \( i\) is defined as the sum of the assigned tasks' weights, \( L_i \). The objective is to minimize the maximum load, \(\min(max(L_i))\), across all processors and return the corresponding load distribution. AMReX uses a distribution map to describe load distributions. A distribution map is a vector \( N \) items long. Each value in the vector stores the MPI rank to which the corresponding box is assigned. A commonly used metric to quantify the quality of a load balancing solution is the `efficiency', $\epsilon$. It is defined as the average load per rank (the optimal load balance, regardless of whether this assembly is possible) over the maximum load across all ranks (the maximum weight, which defines the expected time-to-solution), as given in Equation \ref{eqn:eff}:

\begin{equation}
\label{eqn:eff}
    \epsilon = \frac{{\sum_{j}^N(w_j)}/{P}}{max(L_i)}
\end{equation}

Mathematically, this dynamic load balancing problem is a variation of the `partition problem' \cite{chopra1993partition} or more specifically, the `multiway-number partitioning problem' \cite{korf2009multi}. These problems are NP-hard \cite{hochba1997approximation} and fall within a class of optimization problems focusing on efficient scheduling. Solving these optimization problems directly is expensive, time-consuming, and ineffective as real-time dynamic load balancers inside HPC simulations. So, modern distributed memory problems use a small variety of fast approximation algorithms to achieve reasonable solutions quickly in situ. The most common are Knapsack, or the greedy algorithm \cite{akccay2007greedy}, the largest differencing method or Karmarkar-Karp \cite{korf1998complete}, and space-filling curve (SFC) algorithms \cite{breinholt1998algorithm}.

This investigation focuses on the currently used AMReX algorithms, Knapsack and SFC, as well as three additional algorithms: a brute force solver that is studied to determine its effective use cases in algorithmic analysis, an SFC algorithm using the painter's partition algorithm to determine bisection rather than AMReX's percentage-tracking strategy, and a combination Knapsack + SFC algorithm to investigate whether the benefit of both strategies can be combined. Each of these algorithms is described in detail below:

\subsection{Brute Force}
The brute force algorithm calculates the cost of every possible distribution of the work, tracking the optimal solution as it goes. This approach guarantees that the global optimum is found. However, the required testing space grows exponentially as the problem size increases, which limits it's usefulness as a dynamic load balancer. However, a brute force solver can still be useful to help evaluate the outcomes of other algorithms.

The brute force algorithm designed for this application utilizes a counting methodology. If we have \( N \) boxes assigned to \( P \) processors, all possible combinations of the distribution map can be found by counting in base \( P \) until we reach a number longer than \( N \) digits. For example, if we have \( 4 \) boxes to be assigned to \( 2 \) MPI ranks, all the distributions are given by counting in binary (base 2) until there are more than \( 4 \) digits: $0000$, $0001$, $0010$, $0011$, $0100$, $0101$, $0110$, $0111$, $1000$, $1001$, $1010$, $1011$, $1100$, $1101$, $1110$, $1111$. These maps are combined with a weights vector to calculate the maximum load and report the optimal solution.

As the number of boxes and buckets increases, the number of possible combinations increases exponentially, as given by Equation \ref{bruteeqn}. This can render a brute-force solution computationally infeasible quite quickly. However, there is a symmetry in the problem: load balancing algorithms often treat processor indexing as interchangeable. For example, 0011 and 1100 can be treated as the same distribution, as the same boxes are grouped together, yielding the same processor weights.

\begin{equation}
\label{bruteeqn}
    Combinations = (processors^{boxes})/2+1
\end{equation}


Recognizing this, brute-force solvers only need to check the first half of distributions, because the second half is a permutation of rank indexes. This improvement will not prevent the brute-force from becoming infeasible as an in-situ algorithm, but it does expand the usefulness of the solver and demonstrates a potential strategy for further improvement: elimination of identical, repeated permutations. Despite the computational limitation, a brute force solver can be useful as a static tool to study the accuracy of the approximate algorithms and may even be applied for very small problems when removing load balancing as a variable is desirable, such as during debugging.

\subsection{Knapsack, or the greedy algorithm}
The greedy approach, referred to here as Knapsack, is a single-pass algorithm that assigns the box with the largest remaining weight to the bin with the smallest total weight \cite{calvin2003average}. First, the boxes are sorted in descending order of their weights. Then, the \( P \) largest weights are each assigned to a different rank. Ranks are sorted by ascending order of their total weights. Then, the algorithm assigns the next heaviest box to the lightest rank, re-sorting ranks after each assignment until all have been assigned.

This greedy approach minimizes the maximum weight after each assignment, making the workload distribution as even as possible throughout the assignment process. With many distributions of weights, this can be a quick strategy that often yields a nearly optimal load balance. AMReX's knapsack algorithm performs an additional step: the largest-weight and smallest-weight processors are examined to see if a box can be re-assigned such that the solution is improved. If so, that box is transferred and the check is repeated until no further improvement can be found. This step helps AMReX's greedy algorithm maintain a high efficiency across a broader variation of simulations.

While the greedy algorithm is effective, in practice, this approach exposes other issues. Some common problems include mappings that trigger expensive redistributions, mappings that put too much memory on a single rank, or failing to account for other dimensions of the code's performance, such as communication costs.

\subsection{Space Filling Curve algorithms}

Space-filling curve (SFC) algorithms attempt to improve the overall performance of a simulation by constraining the problem so that boxes which neighbor each other in the computational domain are put on the same or nearby processors. This can reduces the amount of data that must be transferred during expensive inter-node communication routines and improve the overall simulation time \cite{hilbertsfc}. This physical locality is quantified through a SFC. SFCs convert multi-dimensional space into one-dimensional curves with different characteristics, depending on the conversion. A high-locality SFC, such as Morton or Hilbert, is constructed such that points nearby in three-dimensional space are also nearby along the one-dimensional curve.

\begin{figure}[t]
  \centering
  \begin{subfigure}[b]{0.48\linewidth} 
    \includegraphics[width=\linewidth]{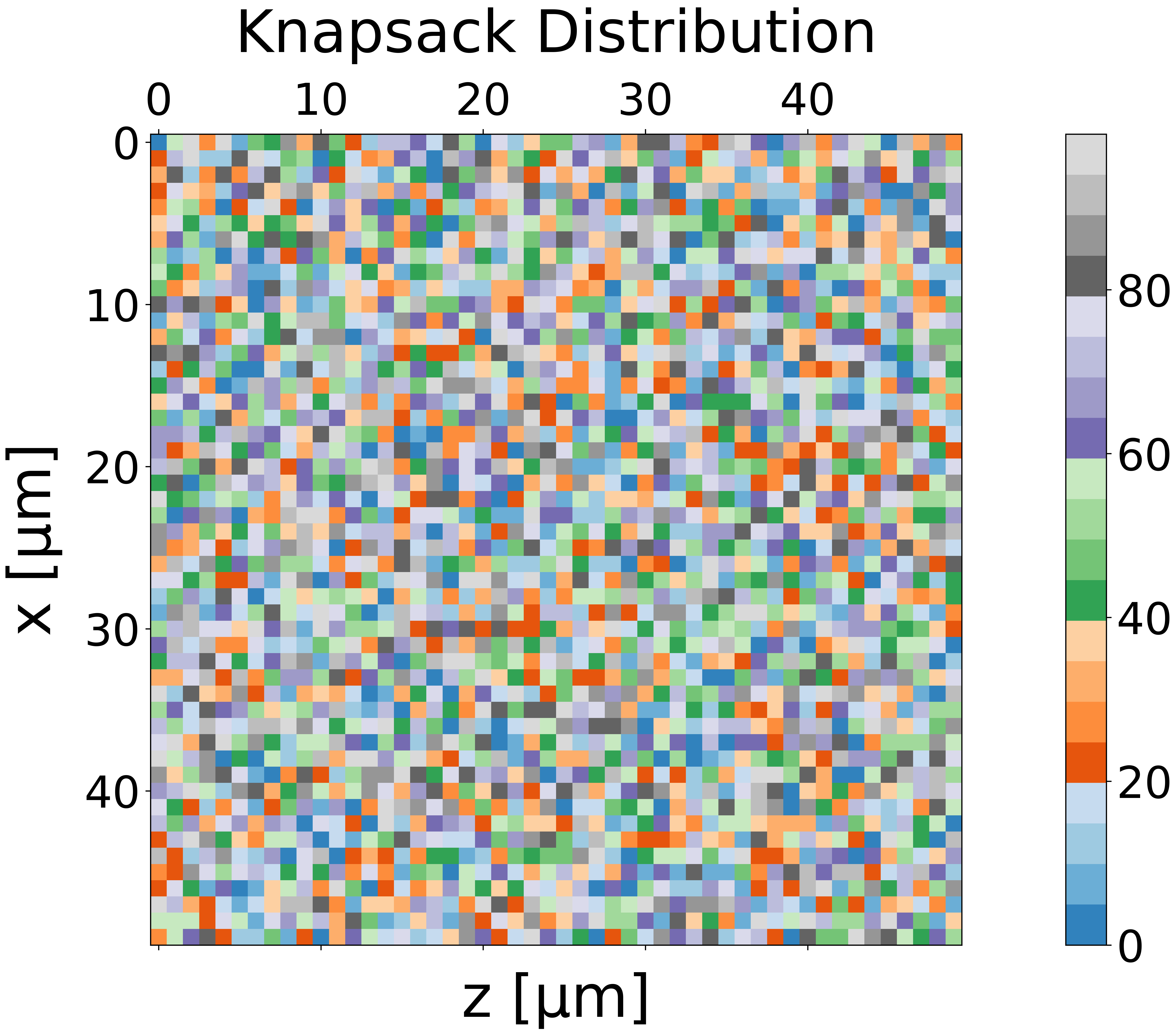}
    \caption{Knapsack load balancing distribution.}
    \label{fig:knapsack}
  \end{subfigure}
  \hfill 
  \begin{subfigure}[b]{0.48\linewidth} 
    \includegraphics[width=\linewidth]{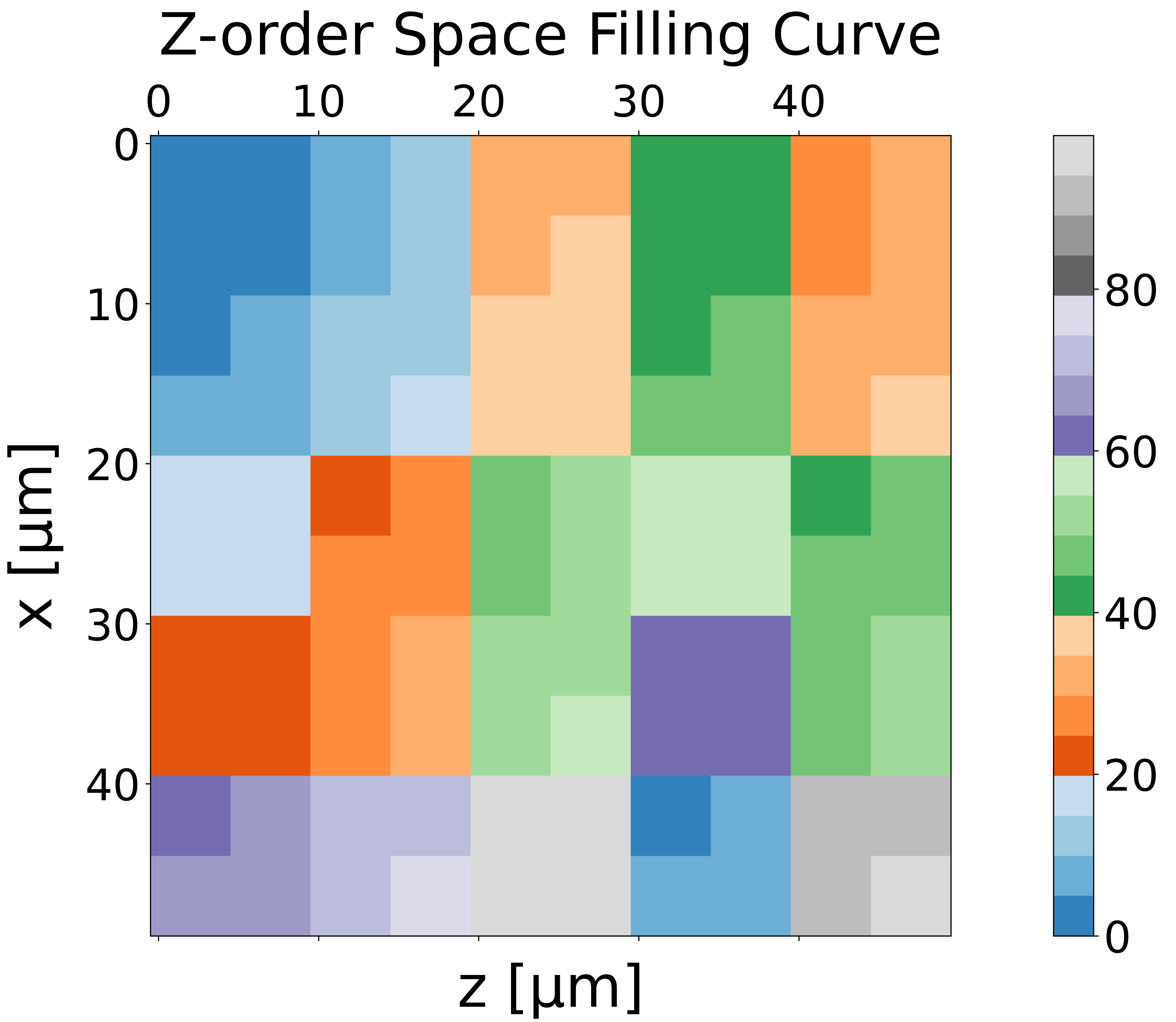}
    \caption{Z-order SFC load balancing distribution.}
    \label{fig:sfc}
  \end{subfigure}
  
  \caption{A comparison of AMReX's Knapsack (a) and SFC (b) domain decompositions after load balancing with an identical set of weights, colored by MPI rank assignment. Knapsack's result distributes the boxes randomly with respect to simulation proximity, while SFC keeps nearby boxes together, which can substantially reduce communication costs across neighboring MPI ranks.}
  \Description{This figure compares AMReX's Knapsack and SFC load balancing distributions. The Knapsack approach appears random, while SFC groups nearby boxes together into similar colored rectangular regions.}
  \label{fig:lbcompare}
\end{figure}

SFC algorithms take advantage of this by converting the multi-dimensional simulation space into an ordered one-dimensional list of boxes. This ordered list keeps boxes that neighbor each other in the simulation close to each other along the sorted list of boxes. If the sorted box list is partitioned into contiguous sections, the partitions will, therefore, contain neighboring boxes. This simultaneous load balances the user-defined weights and substantially reduces communication costs for operations like halo exchanges. The impact of the SFC strategy is illustrated in Figure~\ref{fig:lbcompare}.

AMReX's SFC algorithm uses a Morton (or Z-order) space-filling curve. Once the SFC curve has been applied to create a sorted list of weights, the list is traversed and summed until adding the next weight would exceed the optimal weight (the average weight per processor). If adding the next weight to this processor would keep the total percentage of all weights partitioned equal to or less than the total percentage of all processors so far partitioned, the box is included on that processor, and the next one is begun. Otherwise, the final box is left for the next processor. This percentage-tracking partitioning compensates for under-weighted processors by occasionally making a slightly over-weighted processor when reasonable to do so. This algorithm has improved communication costs in large-scale and GPU-driven AMReX applications, including WarpX's Gordon Bell winning simulations \cite{fedeli2022pushing}, but there is potential for improvement.

\subsection{Painter's Partition algorithm}

Although the SFC algorithm has shown great success in production-scale scientific simulations, the partitioning method is prone to inefficiencies. AMReX's SFC percentage-tracking strategy is designed to create over-loaded processors when a perfect balance cannot be found while traversing the sorted box list. Investigating this inefficiency led to an analogous problem cited around the internet: the painter's partition algorithm. This problem has been used as an interview question for code developer applicants. Still, it provides a fascinating insight into optimally partitioning an ordered, weighted set \cite{PPPGeeks, PPPTakeUForward, PPPApna}.

\begin{figure}[h]
\centering
\includegraphics[width=0.6\linewidth]{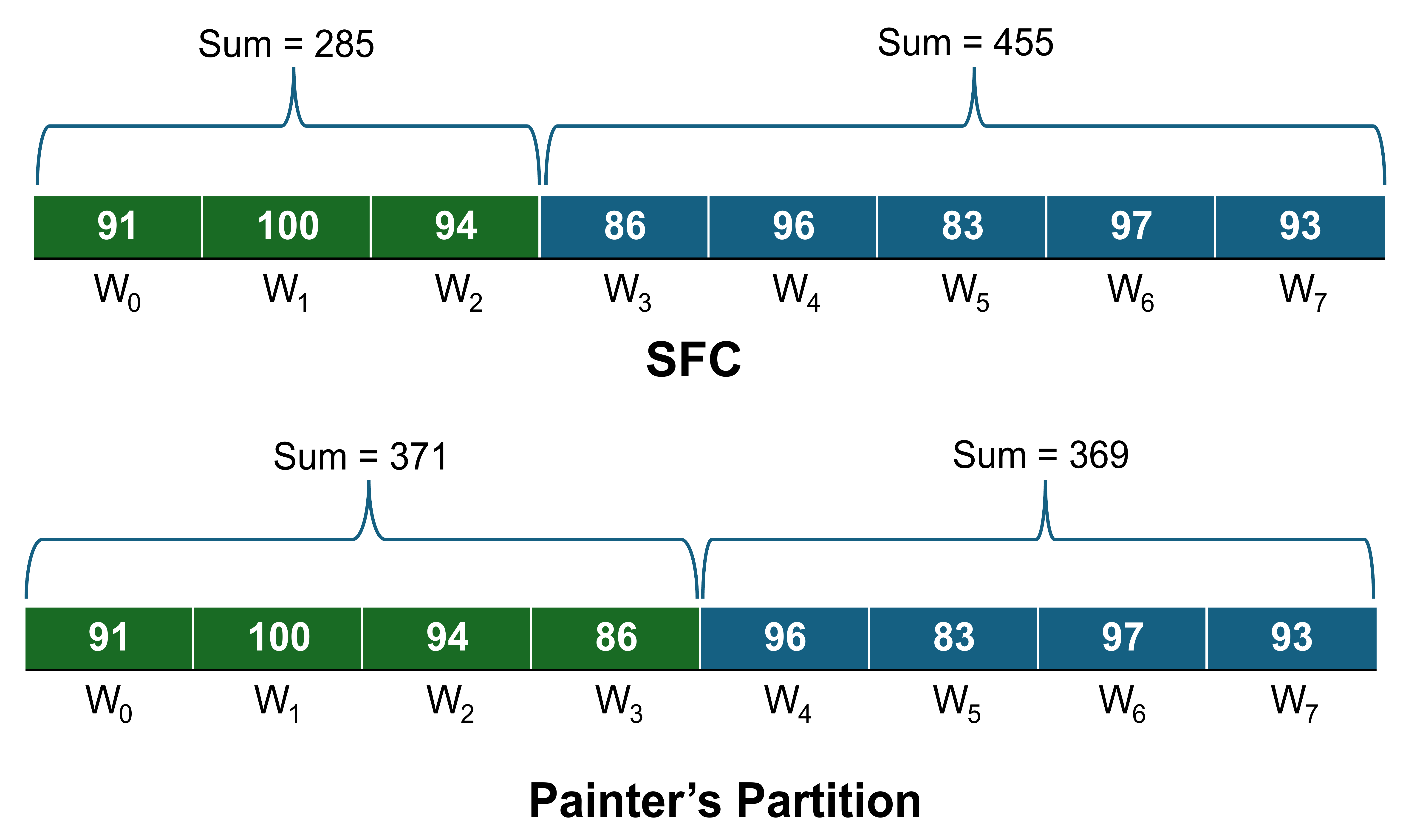}
\caption{A partitioning strategy comparison of AMReX's original SFC algorithm compared to the painter's partition algorithm. In this case, the original SFC algorithm partitions based on the average weight per rank, $370$, and the Painter's SFC algorithm uses a partition size of $371$.}
\Description{Two rows, representing boxes sorted by a space-filling curve, are broken into $8$ sections each, representing boxes, and are numbered by their corresponding weights. The row representing the SFC algorithm is split into $3$ and $5$ boxes, yielding a poor balance, while the Painter's partition row splits into $4$ and $4$, achieving near-perfect balance. }
\label{fig:load_balancing}
\end{figure}

Figure~\ref{fig:load_balancing} demonstrates the difference the painter's partition algorithm can make on the SFC partitioning process. It illustrates workloads represented in as boxes \(w_0, w_1, w_2, w_3, w_4, w_5, w_6, w_7\) with values \(91, 100, 94, 86, 96, 83, 97, 93\), being partitioned across two processors,  green and blue. The SFC algorithm divides the workloads based on the average value ($370$, in this case). The green rank gets \(91, 100, 94\) (sum = $285$) but cannot add \(86\) as that would create a processor with a weight of $371$. So, the blue rank takes the rest, resulting in a substantial imbalance. On the other hand, if an algorithm were able to pick the optimal value for the partition, $371$, it would be able to balance considerably better.  


The Painter's Partition algorithm uses a binary search to achieve this optimal partitioning value. The sorted box list is built identically to AMReX's SFC algorithm. Then, the painter partition performs a binary search across the possible range of target weights: the sum of all weights (the maximum time) and the single largest weight (the minimum time). If the target weight is possible to partition with our fixed number of processors, the target weight is reduced to look for a faster solution. However, if the target weight is too short and more processors are required than are available to split the work, the target weight is increased to look for a valid solution. This is repeated until the ideal target weight is identified. This binary search-driven approach ensures the optimal 1D partitioning is performed.

The Painter's Partition algorithm is detailed in Algorithm~\ref{algo}. `paintersSearch' performs the binary search between `max(weights)' and `sum(weights)', tracked over time by `l' and `h', respectively. It calls `isPartitionPossible' after each adjustment, which attempt to partition the weights with the target number of processors and returns whether it was successful. Once the optimal partition size, `res', has been found, it is returned to construct the optimal distribution map.

\begin{algorithm}[t]
\caption{Painter's Partition Algorithm}
\small
\SetAlgoLined
\DontPrintSemicolon
\KwData{$number\_of\_processors, weights$}
\KwResult{$res$: Target size for the processor distribution}

\SetKwFunction{FMain}{paintersSearch}
\SetKwProg{Fn}{Function}{:}{}
\Fn{\FMain{$weights, no\_of\_boxes, number\_of\_processors$}}{
    $ h \longleftarrow sum(weights) $ \;
    $ l \longleftarrow max(weights) $ \;
    
    \While{$l < h$}{  
        $mid \longleftarrow l+(h-l)/2$ \;
        
        \If{isPartitionPossible(weights, no\_of\_boxes, number\_of\_processors, mid)}{
            $ res \longleftarrow mid $ \;
            $ h \longleftarrow mid - 1 $ \;
        }  
        \Else{
            $ l \longleftarrow mid + 1 $ \;
        }
    }
    \Return $ res $ \;
}

    
            

\label{algo}
\end{algorithm}


\subsection{Combination Knapsack + SFC algorithm}

We also investigated an algorithmic combination of the Knapsack and SFC algorithms to try to capture the key features of both algorithms. This algorithmic design is based on the observation that inter-node communication dominates communication costs. It attempts to use the SFC algorithm to reduce communication costs, followed by the Knapsack algorithm to refine the solution and further improve the balance of the weighted work as well. This novel combination has been developed and tested as part of our investigation.

First, SFC maps the multidimensional data across a node number of partitions, maximizing geometric locality and minimizing expensive inter-node communication. This results in partitions with a large number of neighboring boxes that need to be further partitioned into individual ranks. That is performed by running a Knapsack separately on each node's assigned boxes to distribute optimally across ranks. This may help fine-tune each node's load balance and both improve local calculation balance while also respecting the globally reduced inter-node communication.

Algorithm~\ref{alg:sfc_knapsack_combined} shows the details of this methodology. The SFC algorithm, `RUN SFC', first distributes across `nnodes' partitions. Then, separate `local weights' vectors are built corresponding to each pre-partitioned node. These `local weights' are run through Knapsack, `RUN KNAPSACK', to optimize the result locally, before assembling the final full result and returning it to the application. 

We investigate using both AMReX's original SFC algorithm as well as the painter's partition based SFC algorithm for this combination algorithm. Results are reported for both algorithms independently.

\begin{algorithm}[h]
\caption{Combined SFC \& Knapsack Algorithm}
\label{alg:sfc_knapsack_combined}
\SetAlgoLined
\DontPrintSemicolon
\KwData{boxes, weights, nnodes, ranks\_per\_node}
\KwResult{final\_map: array of global ranks for each box, \\
Combined\_Algorithm\_eff: Final combined efficiency}

\SetKwFunction{FMain}{CombinedAlgo}
\SetKwFunction{FDist}{RUN SFC}
\SetKwFunction{FKnapsack}{RUN KNAPSACK}

\SetKwProg{Fn}{Function}{:}{end}

\SetKwProg{Fn}{Function}{:}{}
\Fn{\FMain{$boxes, wgts, nnodes, ranks\_per\_node$}}{
    
    $total\_weight \leftarrow \sum wgts$ \\
    $vol\_per\_team \leftarrow total\_weight / nnodes$ \\
    
    $sfc\_map [node][assigned box] \leftarrow$ \FDist($tokens, wgts, nnodes, vol\_per\_team$) \\
    
    Compute node weights and $max\_node\_weight$ \\

    $final\_map \leftarrow$ array of size $N$  \\
    $max\_weight\_across\_ranks \leftarrow 0$ \\

    \For{$node \leftarrow 0$ \KwTo $nnodes-1$}{
       $local\_boxes \leftarrow sfc\_map[node]$ \\
       Extract $local\_wgts$ for these boxes \\
       $(knapsack\_res) \leftarrow$ \FKnapsack($local\_wgts, ranks\_per\_node$) \\
       
%
%
%
%
    }

    $Combined\_Algorithm\_eff \leftarrow \frac{total\_weight}{ranks\_per\_node \times nnodes \times max\_w\_across\_ranks}$ \\
    \Return $(final\_map,Combined\_Algorithm\_eff )$
} 
\textbf{end Function}

\end{algorithm}

\section{Methodology}
We evaluated the performance of these load-balancing algorithms on the Perlmutter supercomputer at the Lawrence Berkeley National Laboratory \cite{NERSCperl}. Performing a full statistical analysis of the approximation algorithms at scale would have required a prohibitively large number of full-scale simulations. Instead, weight vectors were generated through a random number generator. We chose a normal distribution for this study as it is a realistic distribution that is amongst the easiest to interpret, discuss and extrapolate to other distributions applicable to readers' use cases.

Two analyses were performed during this study. First, the brute force solver was studied to determine under what conditions it could feasibly be used as a dynamic load balancer and a tool to analyze other load balancing algorithms. The brute force solver was ran at increasing sizes, both in a number of boxes and a number of processors.

The second analysis is a comparison of the approximation algorithms presented in section 2: the knapsack, SFC, painter's, and combination algorithms. This analysis seeks to understand when each algorithm should be considered for use and whether they run fast enough to be used for dynamic load balancing. This study uses randomly generated a distributions of weights and runs this set of weights through each algorithm separately. This is repeated \( 250 \) times with new weight distribution to draw statistical conclusions about applying these algorithms to any normal distribution of weights. The time-to-solution and efficiency of each algorithm are output and analyzed.

AMReX's load-balancing algorithms were extracted from the AMReX framework and set up to run independently. However, this analysis requires a domain to which a space-filling curve can be applied. This domain is constructed and decomposed using AMReX's BoxArray class. A (256 x 256 x 256) domain is constructed and cut into boxes with a given "max grid size" to obtain a decomposed domain consisting of the target number of boxes. For example, a (128 x 128 x 128) "max grid size" will cut the domain into $8$ boxes, $2$ in each geometric direction. This BoxArray defines the location of each box in the domain and is used as input to the space-filling curve to generate ordered weights prior to partitioning.

AMReX's dynamic load-balancing methodology runs the algorithms serially. Specifically, AMReX collects all weights on a single rank, runs the user-chosen load-balancing algorithm on that rank, and communicates the result to all other ranks before redistributing the problem, if desired. This strategy is used because (1) the approximation algorithms are currently so fast that parallelizing them is unnecessary; and (2) to avoid errors from different ranks calculating different mappings due to effects like rounding errors. This study seeks to investigate the relative time-to-solution of the approximation algorithms for AMReX and similar applications. So, while there may be parallelization opportunities, this study restricts itself to serial studies of the approximation algorithms.

This approximation algorithm analysis uses three different normal distributions to study the effect of a variation in  weight distributions on overall efficiency. These three distributions have a mean of \( 100000 \) and vary the standard deviation to increase the variability of weights. We analyze a normal distribution with a "small" standard deviation of \( 250 \), a "medium" standard deviation of \( 4523 \), and a "large" standard deviation of \( 25231 \).

The analysis of the approximation algorithms was ran based on the range of use cases of AMReX on Perlmutter's GPU partition. \(4\) ranks-per-node is used in the combination algorithms to match the number of GPUs present on a GPU node of Perlmutter. Currently, the majority of HPC algorithms use one rank-per-GPU to run their jobs, so the number of nodes is varied from 1 to $512$, by factors of $2$. $512$ represents a one-third scale job on Perlmutter, which has $1792$ GPU nodes. Finally, the number of boxes is varied by testing 4, 8, and 16 boxes-per-rank for each node size. This aligns with AMReX's typical use cases: at least $4$ boxes-per-rank are needed to provide enough options to allow a useful partitioning, but using more than $16$ boxes-per-rank starts to cause performance issues due to excessive memory fragmentation. Appendix A provides the location of our reproduction artifact, which includes the code, input files, results, analysis notebooks, and detailed build instructions.


\section{Results and Discussion}
\subsection{Brute Force Solver}
\begin{figure}[t]
  \centering
  \begin{subfigure}[b]{0.48\linewidth} 
    \includegraphics[width=\linewidth,keepaspectratio]{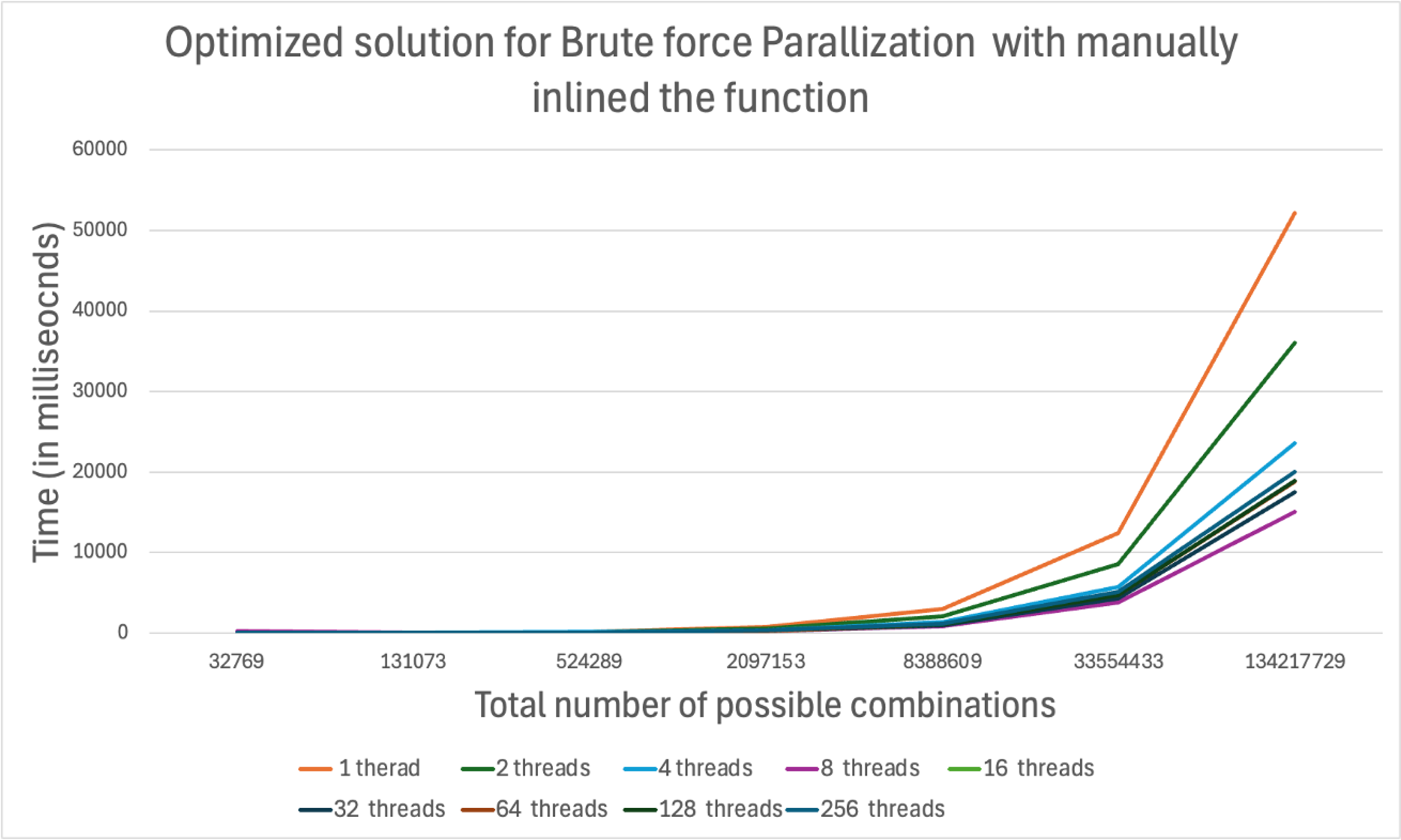}
    \caption{Time-to-solution for the brute force solver, including OpenMP parallelization}
      \Description{A showing the increase in time to solution with respect to the number of combinations that much be solved and the number of OpenMP threads used to solve the problem. All OpenMP curves show an exponential growth in time to solution as the number of combinations increases.}
    \label{fig:parallization_bruteForce}
  \end{subfigure}
  \hfill 
  \begin{subfigure}[b]{0.48\linewidth} 
    \includegraphics[width=\linewidth,keepaspectratio]{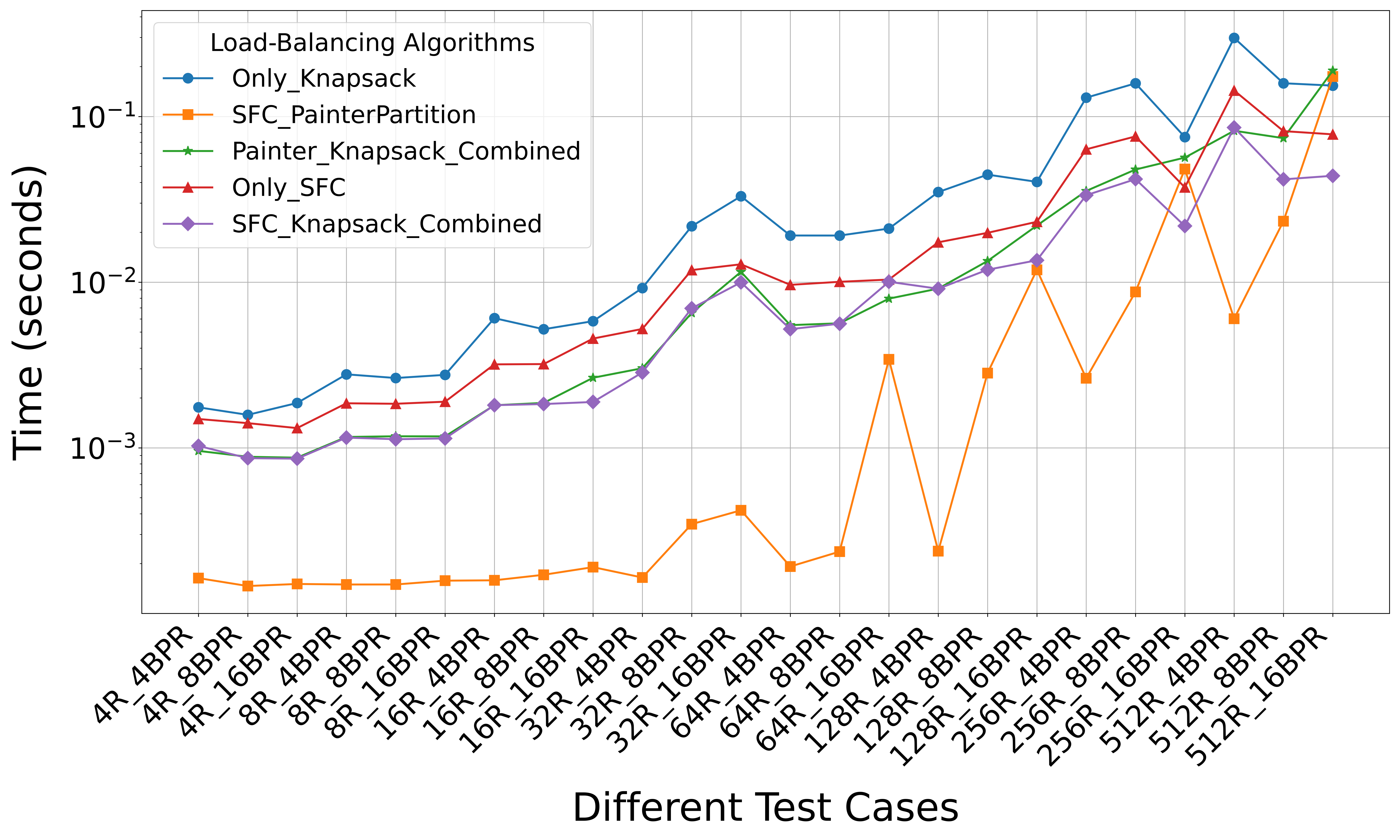}
    \caption{Average time to solution for the approximation algorithms, plotted on a log scale.}
   \Description{Plot of time to solution for the approximation algorithms. The knapsack, SFC, and combination algorithms show similar, overlapping lines falling between $1$ and $0.001$ seconds across all test cases. The painter's partition line is substantially lower, staying below $0.001$ seconds until around 128 ranks, when it increases to match the other algorithms.}
    \label{fig:time_best}
  \end{subfigure}
  
  \caption{Average time to solution for the tested algorithms. Test cases are named X(R)\_Y(BPR), representing X ranks and Y boxes-per-rank. This plot is of the lowest standard deviation cases. Medium and high standard deviation results are extremely similar and therefore are not included. Those plots can be found in the reproducibility artifact.}
  \label{fig:lowstddev}
\end{figure}

In this first study, the brute force solver is timed to determine when it can be practically used. First, it was optimized to improve scalability and extend its usefulness. Key functions were manually in-lined to reduce overhead created by the extremely large number of repeated calls. The main counting loop was parallelized using OpenMP threading, which allowed up to 256 simultaneous distributions to be examined on Perlmutter. However, given the exponential growth in complexity with problem size, as described in Equation \ref{bruteeqn}, this only slightly increased the range of applicability, as shown in Figure~\ref{fig:parallization_bruteForce}.

A reasonable estimate for the maximum time for a non-restrictive dynamic load balance in smaller AMReX applications is around $20$ seconds. As shown in Figure~\ref{fig:parallization_bruteForce}, this means using at least $8$ threads and solving problems with less than $134$ million combinations. This corresponds to load balancing problems of $4$ ranks and $3$ boxes per rank or $2$ ranks and $14$ boxes per rank. These are incredibly small problems, demonstrating conclusively how infeasible a brute force solver is in the vast majority of HPC simulations. However, the brute force solver can still be used independently where time-to-solution is not critical, such as when analyzing the efficiency of other load balancing strategies.


\subsection{Approximation Algorithm Analysis}

\begin{figure}[t]
  \centering
  \begin{subfigure}[b]{0.48\linewidth} 
    \includegraphics[width=\linewidth]{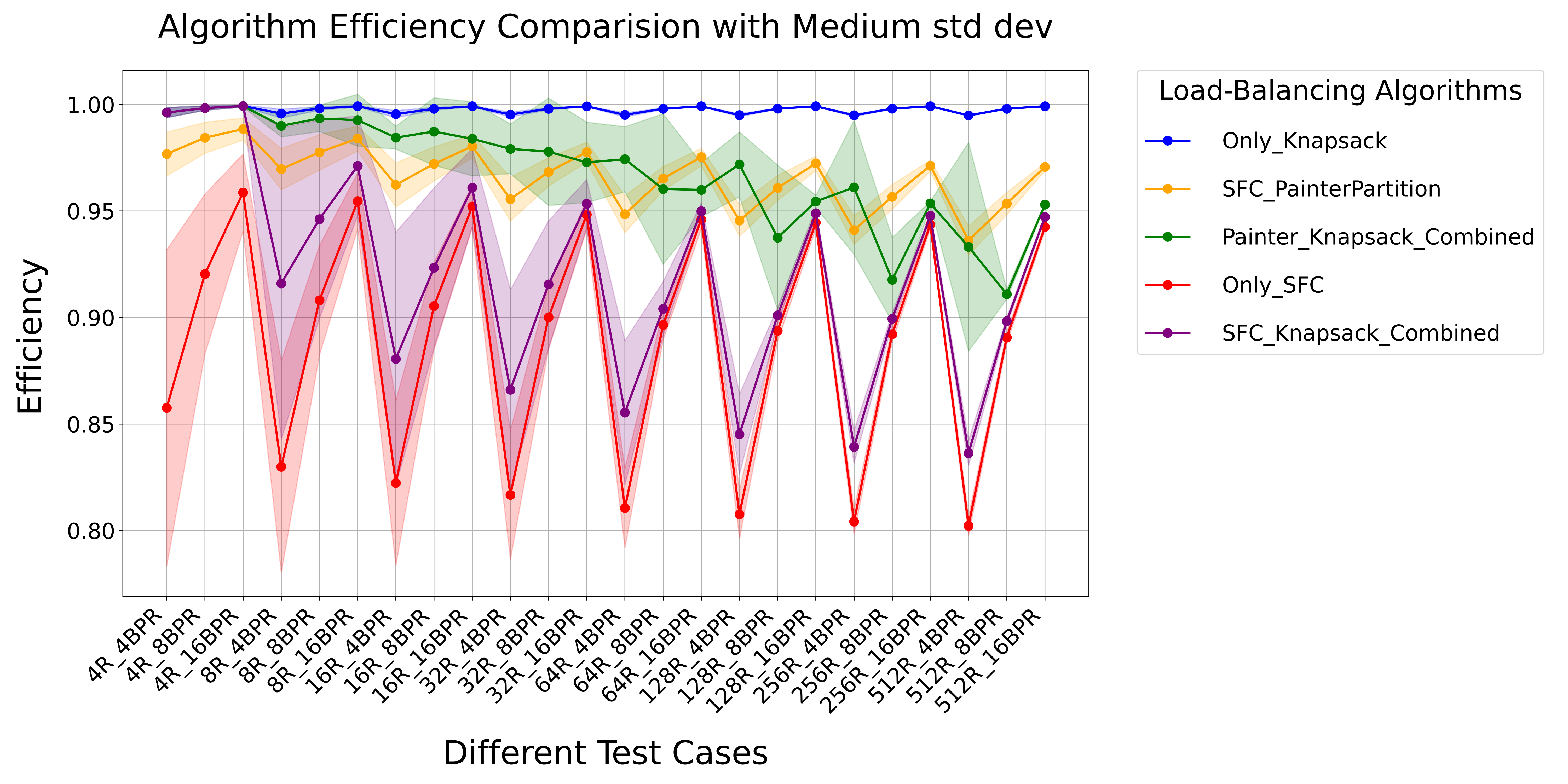}
    \caption{Medium standard deviation $--4523$}
      \Description{Knapsack remains the most optimal. SFC and the original SFC combination also show similar results compared to a low std. dev., although the original SFC combination is starting to appear slightly better than SFC. Painters and Painters combination algorithms show decreasing efficiency as problem size increases. By 512 ranks, they are very close to the SFC algorithms, ~ $0.92$.}
    \label{fig:efficiency_avg}
  \end{subfigure}
  \hfill 
  \begin{subfigure}[b]{0.48\linewidth} 
    \includegraphics[width=\linewidth]{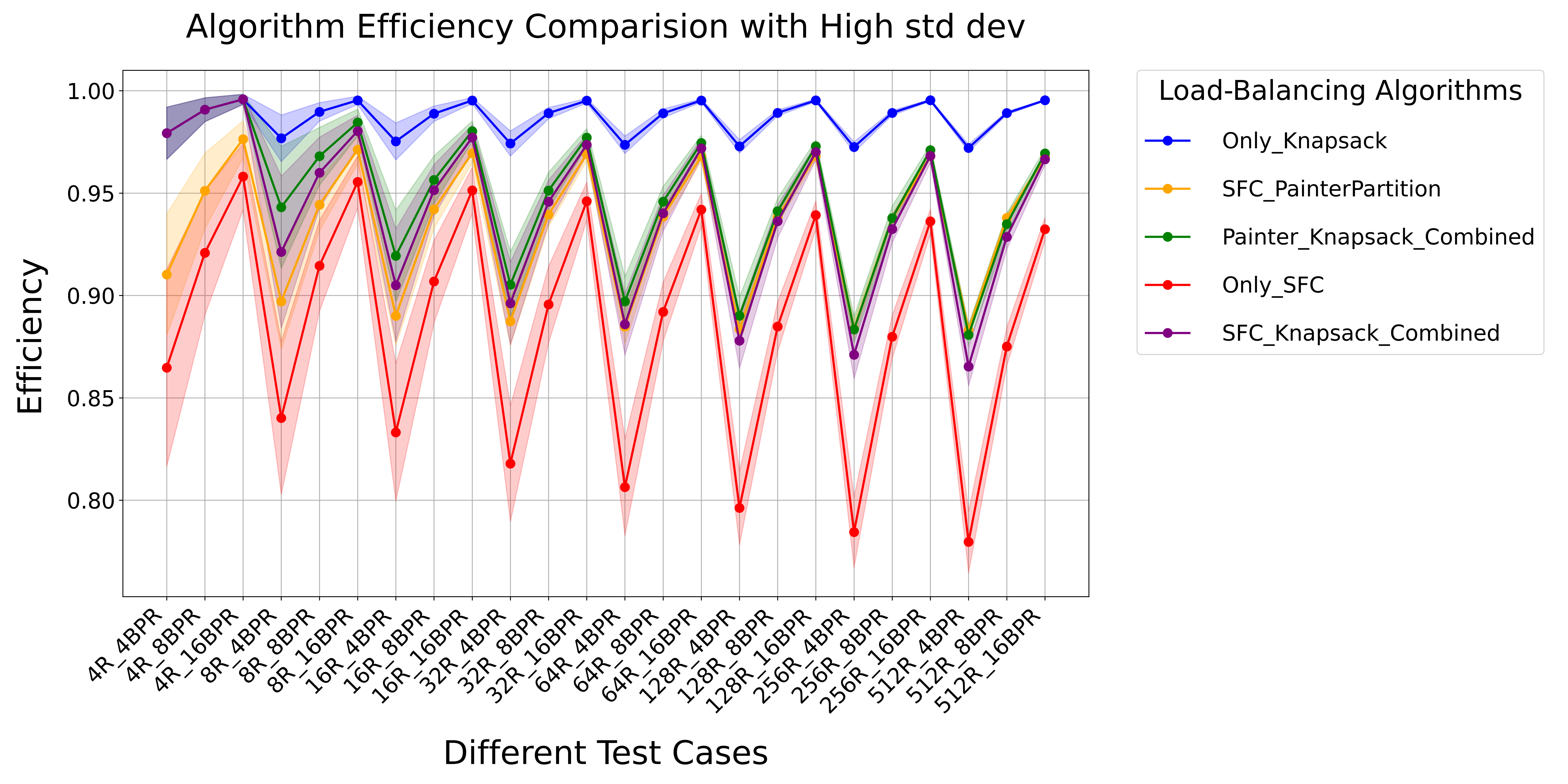}
    \caption{High standard deviation $--25232$}
  \Description{All algorithms again show a definite pattern of improved performance when boxes-per-rank is increased. Knapsack is well above the other algorithms (greater than $0.97$) followed by painters and the combination algorithms, which appear to yield similar results, but lower than in the other plots.. SFC is the lowest, but has the same efficiency range as the other plots $0.8$ to $0.95$.}
    \label{fig:efficiency_worst}
  \end{subfigure}
  
  \caption{Average efficiency comparison between partitioning algorithms for a weight distribution with a medium standard deviation in (a) and with high standard deviation in (b).}
  \label{fig:highstddev}
\end{figure}

As a first step in evaluating the approximation algorithms, we need to confirm these algorithms are fast enough to be used as dynamic load balancers. The average time to solution for each algorithm for sizes up to $512$ ranks is given in Figure~\ref{fig:time_best}:


This plot shows that all approximation algorithms complete in under a second, more than fast enough to be efficient, practical real-time dynamic load balancers. In this plot, we also see that the painter's partition algorithm is substantially faster than the other approximation algorithms at small scales. We believe this is due to our painter's algorithm not including some of the expensive intermediate work that required to map the solution to the simulation. However, the specific cause of this has not been investigated, as this does not impact the conclusion that all of these algorithms are more than fast enough to be used as dynamic load balancers. The next step is to compare these algorithms' efficiencies. The efficiency plots, Figures \ref{fig:lowstddev} and \ref{fig:highstddev}, plot the average efficiency over the $250$ runs with increasing problem size, as well as the standard deviation of the resulting efficiencies in colored shading.

\begin{figure}[t]
  \centering
  \begin{subfigure}[b]{0.48\linewidth} 
    \includegraphics[width=\linewidth]{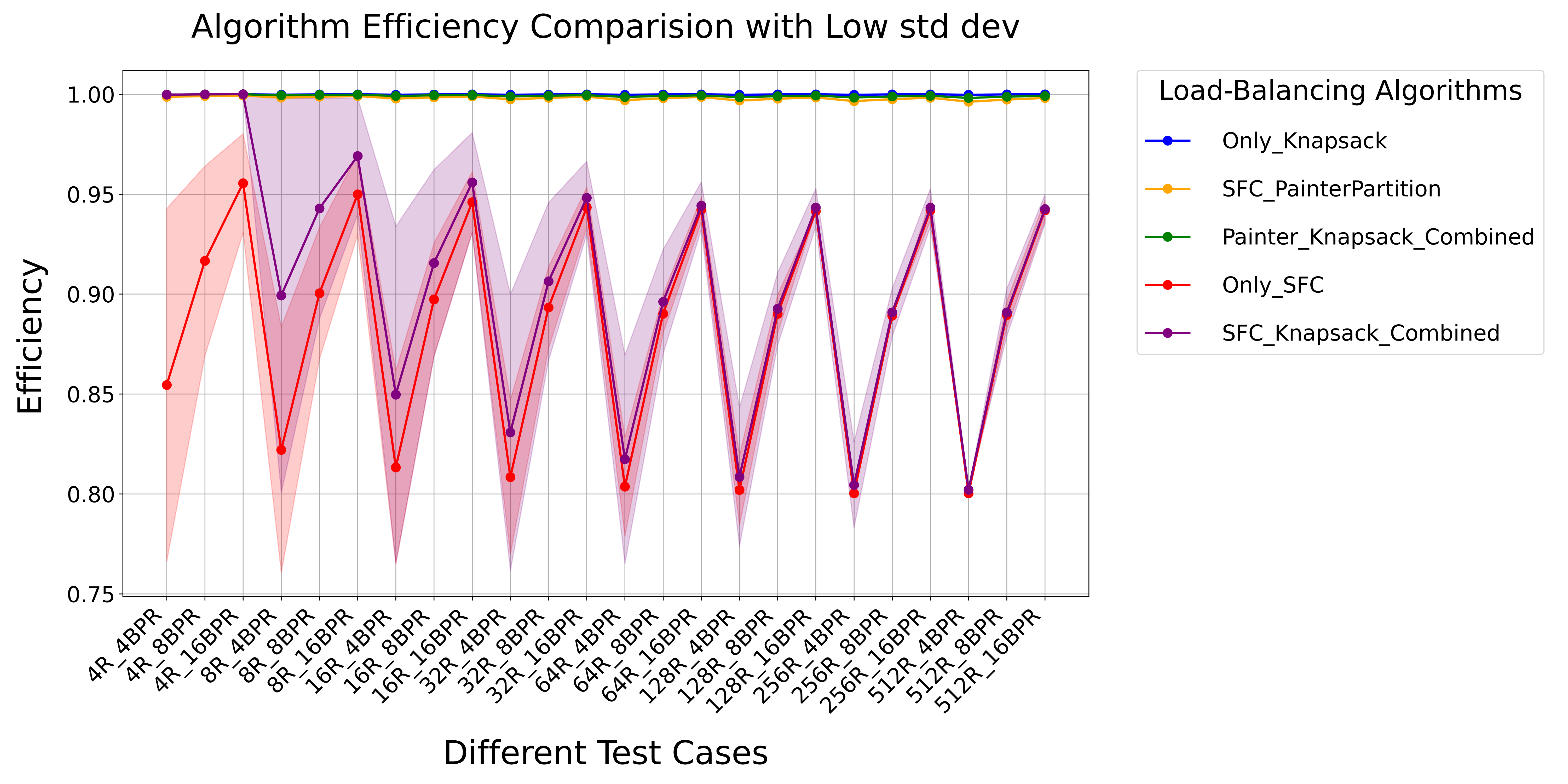}
    \caption{Complete small std. dev. results}
      \Description{Plot of the efficiency for all algorithms over time. Knapsack, painter's and painter's combined yielded near optimal, $1.0$, efficiences across all tests. SFC and original SFC combined yield similar times, with combined being slightly better, between $0.8$ and $0.95$ across all cases. SFC and SFC combined have higher effiencies with higher boxes-per-rank.}
    \label{fig:efficiency_best}
  \end{subfigure}
  \hfill 
  \begin{subfigure}[b]{0.48\linewidth} 
    \includegraphics[width=\linewidth]{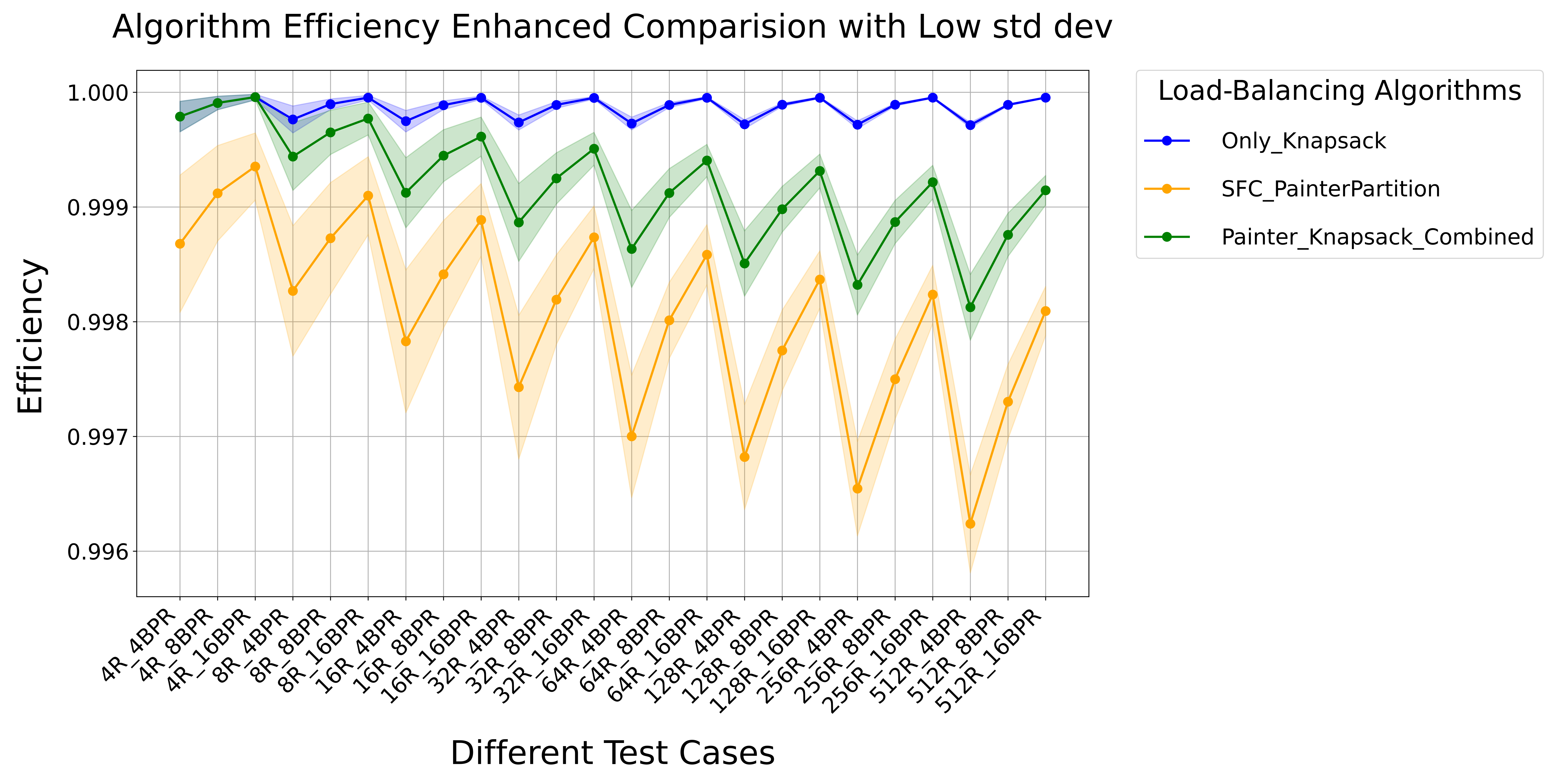}
    \caption{Closeup of near-optimal results}
   \Description{Plot of the efficiency for all algorithms over time. Shows knapsack as the optimal balance, followed by painters and painter's combination algorithms, all very nearly optimal. All algorithms show slight improvement with boxes-per-rank.}
    \label{fig:efficiency_best_zoom}
  \end{subfigure}
  
  \caption{Average efficiency comparison between algorithms for a weight distribution with a small standard deviation $--250$ (a). A close-up of Knapsack, painters, and painter's combination algorithms at nearly perfect efficiency in sub-figure (b). The standard deviation of the results is shown by corresponding color shading.}
  \label{fig:lowstddev}
\end{figure}

First, as a matter of confirmation, these efficiency plots show that the combination algorithms are equivalent to the Knapsack at $4$ ranks, or $1$ node. This is expected, as the SFC step in the combination algorithms should perform no partitioning when solving distributions with only $1$ node. Starting from the investigations with lowest standard deviation, Figure \ref{fig:efficiency_best}, we can see that the Knapsack, painter and painter-based combination algorithms yield near perfect partitioning, regardless of problem size. The original SFC and original SFC+Knapsack algorithms show a less optimal result that is improved with a larger number of boxes per rank. This is a typical behavior for load-balancing algorithms: more boxes provide more options to load balance with, leading to better solutions. However, in this case where weights are very similar, this is because putting one extra box on a rank is significantly less impactful when there are an average of $16$ boxes per rank than an average of $4$ per rank.

The painter's partition methodologies far outperform the original SFC partitioning strategies. From Figure \ref{fig:efficiency_best_zoom}, the combination algorithms consistently equal or at least slightly outperform their equivalent SFC-only algorithms for all problem sizes. This suggests combination algorithms may be viable for a variety of simulations, however, whether the effect of the knapsack step on SFC's communication improvement effects remains unknown and can only be verified by testing on real applications.

Additional insights can be found by comparing these results to the results for a medium standard deviation and a large standard deviation weight distribution, given in Figure~\ref{fig:highstddev}. As the variety of box weights increases, the painter's algorithms efficiency collapses towards the original SFC algorithms efficiencies, suggesting these algorithm types are similarly efficient when box sizes are more varied and more challenging, weight-based distributions are required to find optimal solutions. 

The decrease in painter's algorithmic efficiency can be clearly seen when using the medium standard deviation in Figure~\ref{fig:efficiency_avg}, which shows a collapse towards the SFC efficiencies as problem size is increased. Therefore, algorithmic selection likely has less impact during highly complex simulations, such as those with large amounts of localized mesh refinement. More modern complex algorithmic approaches may be required to obtain higher partitioning efficiencies in complex, multi-physics simulations.

The original SFC strategy is very robust, maintaining an efficiency of between roughly $0.8$ and $0.95$ across all tested sizes and distributions. This may make SFC-based algorithms ideal for situations in which sudden load imbalances are prevalent, such as shock waves development. However, all of the painter's algorithms maintain at least a slight lead over the SFC algorithms for the entire range of differences, suggesting painter's partitioning should be used wherever possible. The only exception appears at the largest standard deviation: we begin to see the original SFC combination algorithm start to achieve efficiencies similar to the Painter's algorithms. Future studies should explore if there are any distributions in which percentage-tracking combination algorithms can outperform the painter's partitioning algorithm.

\section{Conclusion}
This paper investigates various load-balancing algorithms suitable for dynamic load balancing in domain decomposition-based high-performance computing (HPC) simulations. A brute force algorithm was developed, optimized, and examined. The results indicated that while this algorithm can serve as a useful tool for evaluating load-balancing strategies, it is only practical for the smallest use cases as a dynamic load balancer. Among the tested approximation algorithms, the painter's partition-based algorithms outperform the original SFC-based strategies across all tested cases and should be considered wherever maximum performance is the primary concern. Additionally, combination algorithms outperform their single-algorithm counterparts and should be evaluated for potential use in production-scale simulations. However, the difference in algorithms reduces as the weights being balance become more varied, suggesting more complex algorithms may need to be developed to further optimize modern simulations with a wide-variety of physics, particle interactions and other complex flow considerations.

\section{Future Work}
This research has shown substantial improvement is possible for current-generation load-balancing strategies. This has led to interest in further research in optimizing load-balancing strategies in AMReX applications. Areas of interest for future work include: (1) further refinement of current algorithms by investigating different space-filling curves, such as Hilbert \cite{hilbertvmorton, hilbertsfc}, (2) testing these new algorithms in an AMReX application to study the true impact on communication performance (3) expanding to uniform-machine scheduling \cite{uniformmachine} techniques to utilize heterogeneous architectures efficiently, such as cloud resources and (4) whether direct mathematical optimization strategies can be effective as iterative load balancers if implemented asynchronously to the simulation, such as on a separate thread or a separate accelerator, such as a quantum computer \cite{SC24quantum}.

\begin{acks}
This research used resources of the National Energy Research Scientific Computing Center (NERSC), a Department of Energy Office of Science User Facility.
\end{acks}



\appendix
\section{Reproducibility Artifact}
This paper includes a reproducibility artifact conforming with ACM's recommendations. It can be found at \url{https://github.com/amitashnanda/ACM_PEARC_2025_Paper_Artifact.git}
\end{document}